# Optically resolving the dynamic walking of a plasmonic walker couple

*Maximilian J. Urban[1], Chao Zhou[1], Xiaoyang Duan[1], and Na Liu[1,2]\**

[1]Max Planck Institute for Intelligent Systems, Heisenbergstrasse 3, D-70569 Stuttgart, Germany

[2]Kirchhoff Institute for Physics, University of Heidelberg, Im Neuenheimer Feld 227, D-69120 Heidelberg, Germany

ABSTRACT:

Deterministic placement and dynamic manipulation of individual plasmonic nanoparticles with nanoscale precision feature an important step towards active nanoplasmonic devices with prescribed levels of performance and functionalities at optical frequencies. In this Letter, we demonstrate a plasmonic walker couple system, in which two gold nanorod walkers can independently or simultaneously perform stepwise walking powered by DNA hybridization along the same DNA origami track. We utilize optical spectroscopy to resolve such dynamic walking with nanoscale steps well below the optical diffraction limit. We also show that the number of walkers and the optical response of the system can be correlated. Our studies exemplify the power of plasmonics, when integrated with DNA nanotechnology for realization of advanced artificial nanomachinery with tailored optical functionalities.

In the cellular world, materials and information throughout the cell are often transported by a myriad of biological motors, which can carry out nanometer steps along protein tracks[1]. To date, the understanding about how individual motors execute directed movement and how multiple motors work collectively has remained incomplete[2-5]. Nevertheless, this has not hindered but rather fostered a strong incentive to harness artificial nanomachines for performing functions similar to their biological analogs[6]. Among diverse opportunities, DNA walkers represent one of the most successful attempts toward this aim[7]. Over the last decade, miscellaneous DNA walkers, which can perform controlled movement[8-19], transport cargo[20-21] and direct chemical reactions[22], have been accomplished. So far, research efforts on DNA walkers have mostly focused on individual walker behavior. However, implementation of multiple artificial walkers that can walk both individually and collectively on the same track, has not been achieved.

By combining recent advances in nanoplasmonics and DNA nanotechnology[23-29], in this Letter, we demonstrate a novel dynamic plasmonic walker couple, composed of two gold nanorod (AuNR) walkers. We utilize optical spectroscopy to resolve the dynamic stepwise walking of the walker couple powered by DNA hybridization along a shared DNA origami track. We first show that the two walkers can individually execute nanoscale steps directionally and progressively, demonstrating the independent control of each walker. Subsequently, we show that the two walkers can perform simultaneous walking, working together on the same track.

A sensitive plasmonic coupling scheme[30] is introduced for *in situ* optically monitoring the dynamic walking of the two walkers with steps well below the diffraction limit. The two walkers and a stator (an immobilized AuNR) are assembled on the two opposite surfaces of an origami template (see Figure 1a). The two walkers (red and green AuNRs) optically interact with the stator (grey AuNR) simultaneously. In particular, each walker and the stator constitute a chiral geometry[31-34]. The chiroptical response of the entire system is jointly determined by the positions of both walkers relative to the stator. This renders optical discrimination of the walking directions and the individual steps of the two walkers possible. Finally, to examine the possibility of optically determining the walker number, one of the walkers is dissociated off the track. A clear spectral shift is observed. The experimental results show an overall good agreement with the theoretical calculations. Our study exemplifies the power of plasmonics when integrated with DNA nanotechnology for deterministic placement and active manipulation of plasmonic particles with nanoscale precision, as well as for characterization of dynamic DNA nanostructures using optical spectroscopy.

Fig. 1a illustrates the schematic of the plasmonic walker couple system. The nominal dimension of all the AuNRs is 35 nm×10 nm. Double-layer DNA origami[35, 36] is utilized to achieve a rigid and robust track. The DNA origami (58 nm×42 nm×7 nm) was prepared following a typical annealing process[36] and then purified to remove excess strands (see supporting information). Six rows of footholds (*a-f*) with equal spacing of 7 nm are extended from the origami template. Both of the walkers are fully covered with foot strands of the same sequences, which are partially complementary to the footholds on the track. In particular, each walker resides on two rows of footholds in order to establish a stable configuration, when reaching a certain system state.

The individual walking behavior of the walker couple is first investigated as shown in Fig. 1b. In this case, one walker halts, whereas the other walker carries out stepwise walking. At the initial state, *i.e.*, state I, the red and green walkers are symmetrically positioned at the two ends of the stator. More specifically, the red walker steps on rows *a* and *b*, while the green walker steps on rows *e* and *f*. The entire system is therefore achiral at state I. When the green walker stays stationary at rows *e* and *f*, the red walker can carry out two steps with a step size of 7 nm forward, successively reaching states II and III, following the upper route indicated by the solid arrows (see Fig. 1b). On the other hand, when the red walker stands still at rows *a* and *b*, the green walker can perform two discrete steps toward the red walker, reaching system states IV and V, respectively, following the lower route indicated by the dotted arrows (see Fig. 1b).

When excited by light, the plasmons generated in the walkers and the stator, are coupled through near-field electromagnetic interactions[37-39]. Due to the cross configurations formed by the respective walkers and the stator, the collectively coupled plasmons lead to plasmonic circular dichroism (CD). The resulting CD spectra are highly sensitive to the configuration changes of such a three-dimensional (3D) system[40]. Transmission electron microscopy (TEM) was carried out to access the assembled structures. An overview TEM image of the structures at state I is shown in Fig. 2a, demonstrating a good assembly of the walkers and the stator on the origami. The enlarged TEM image is presented in the black-framed inset of Fig. 2a, where the AuNRs and origami are clearly visible. At state I, the red (green) walker and the stator form a right-handed (left-handed) geometry, therefore rendering the overall state achiral. The CD spectrum at state I was measured using a Jasco-1500 CD spectrometer. The result is shown by a black curve in Fig. 2b. A slight left-handed preference is observable in the CD spectrum. This is likely due to the sample imperfection in the experiment, as any minute assembly deviation may disturb the ideal achiral symmetry, leading to measurable signals resulting from the high sensitivity of CD spectroscopy.

The red walker first advances steps toward the green walker as shown by the upper route in Fig. 1b. Such directional walking is activated by toehold-mediated strand displacement processes[29,41] as illustrated in Fig. 2c. In brief, the red walker programmably attaches (detaches) its feet on (from) the track through hybridization (de-hybridization) with the corresponding foothold rows. Blocking strands (*a*') specific to row *a*, and removal strands (*c*") specific to row *c* are added simultaneously. Dissociation of the red walker's feet from row *a* is initiated by the blocking strands *a*' through strand displacement. Meanwhile, the removal strands *c*" undergo a strand displacement reaction and take the responsibility to make the footholds on row *c* accessible for binding the red walker's. As a result, the red walker imposes one step forward in a rolling fashion by stepping on rows *b* and *c*, reaching system state II. The corresponding TEM image is shown in the red-framed inset of Fig. 2a. At this state, the asymmetry of the right-handed geometry formed by the red walker and the stator decreases with respect to that at state I, whereas the left-handed geometry formed by the green walker and the stator is unchanged (see Fig. 1b). Therefore, the net handedness of the system is left-handed. This is reflected by a characteristic peak-to-dip profile in the CD spectrum as shown by the red curve in Fig. 2b. Even though it arises from a 7 nm displacement of the red walker on the track, the chiroptical response is very distinct and the CD intensity is as large as 40 mdeg. Subsequently, the red walker executes one more step forward by adding blocking strands *b*' and removal strands *d*", reaching system state III. The red walker and the stator constitute a nominally achiral geometry (see Fig. 1b). The chiroptical response of the system therefore mainly originates from the left-handed geometry formed by the green walker and the stator. As shown in Fig. 2b, the CD intensity increases to approximately 70 mdeg. The TEM images of the structures at station III are presented in the green-framed inset of Fig. 2a.

Similarly, the stepwise walking of the green walker can be optically tracked. The red walker resides on rows *a* and *b*, while the green walker carries out two successive steps by addition of corresponding blocking and removal strands. The system reaches states IV and V (see the lower route in Fig. 1b). Our system is designed such, that the stepwise walking of the red and green walkers leads to oppositely handed geometries, which are optically distinguishable. More specifically, they are associated with nearly mirrored CD response, when compared the spectra at states II and III with those at states IV and V. In short, through careful structure designs, both the individual walking directions and nanoscale steps of the two walkers on the same track can be optically resolved using standard CD spectroscopy. Theoretical calculations of the CD spectra were carried out using commercial software COMSOL Multiphysics based on a finite element method. The calculated results can be found in supporting information. Overall, the experimental spectra agree well with the theoretical ones.

Apart from the individual walking of each walker, our system allows for the simultaneous walking of both walkers. To in-situ monitor the dynamic walking processes as well as to directly compare the simultaneous and individual walking behavior, the timed-dependent CD response was recorded at a fixed wavelength of 695 nm, *i.e.*, at the CD peak or dip position. We examine the individual and simultaneous walking routes, coded A and B, respectively, as illustrated in Fig. 3a. Both routes A and B begin with state I, followed by state II. It is associated with an immediate CD intensity increase as shown by both the dashed (route A) and solid (route B) curves in Fig. 3b. Then, routes A and B start to differ. By following route A (see Fig. 3a), the two walkers carry out two individual steps to reach system state IV. First, the red walker halts, and the green walker executes one step toward the red walker. In fact, it introduces a new state, *i.e.*, state VI. This state is also nominally achiral in that the two walkers are symmetrically positioned at the two sides of the stator. A clear CD decrease is observed as shown by the dashed curve in Fig. 3b. Next, the green walker halts, and the red walker carries out one step backward. The system reaches state IV. This is followed by a further CD decrease as shown by the dashed curve in Fig. 3b. As a result, the transition from state II to state IV via route A is associated with two clear CD decrease steps.

Along route B, the two walkers execute simultaneous walking to undergo a direct transition from state II to state IV. The simultaneous walking mechanism is illustrated in Fig. 3c. Blocking strands (*c'* and *f'*) specific to rows *c* and *f*, as well as removal strands (*a"* and *d"*) specific to rows *a* and *d* are added at the same time. Through strand displacement processes, the two walkers walk simultaneously by executing a concurrent movement on the track along the same direction. Consequently, the red walker steps on rows *a* and *b*, while the green walker steps on rows *d* and *e*. The system then reaches state IV. As shown by the solid curve in Fig. 3b, this transition is associated with a single-step, yet large, CD intensity decrease. The overall CD response change via route A is smaller than that via route B, likely due to the influence from the steric hindrance between the two closely spaced walkers at the intermediate state VI. Eventually, both routes A and B end at state V. As shown by the solid and dashed curves in Fig. 3b, the two samples exhibit a similar response with further CD intensity decrease at state V. In short, the walker couple may follow different routes by executing individual or simultaneous walking to reach designated system states. This elucidates a sophisticated control of multiple walkers on the same track, afforded by the high programmability of DNA nanotechnology.

Finally, the correlation between the number of walkers and the chiroptical response of the system is investigated. One of the walkers is dissociated off the track. State V is selected as an exemplary starting state because at this state the two walkers are positioned very close to each other. Therefore, the plasmonic coupling aspect between the two walkers can be examined. The TEM images and the CD spectrum (blue curve) at state V are presented in Fig. 4a. By adding the blocking strands (*a'* and *b'*) specific to rows *a* and *b*, the red walker is dissociated off the track (see the left route in Fig. 4b). The green walker still resides on rows *c* and *d*. The system becomes nominally achiral. As shown by the green curve in Fig. 4a, this results in a nearly featureless CD spectrum with a slight left-handed preference. In contrast, addition of blocking strands (*c'* and *d'*) specific to rows *c* and *d* dissociates the green walker off the track (see the right route in Fig. 4b). In this case, the red walker remains on rows *a* and *b*. As shown by the red curve in Fig. 4a, the resulting CD spectrum exhibits a clear spectral red-shift with a stronger spectral profile, when compared to the blue curve at state V. This is due to the fact that when the red walker is left alone on the track, the transverse coupling between the two walkers vanishes, giving rise to a resonance red-shift according to the plasmon hybridization theory[42]. The chiroptical response becomes more pronounced because when the green walker is off the track, the left-handed preference disappears and the asymmetry of the system turns larger. In this regard, the number of the walkers on the origami track can be optically resolved.

In conclusion, we have demonstrated a plasmonic walker couple system powered by DNA hybridization. We have shown a successful control of two plasmonic walkers, which can independently or simultaneously walk on the same track. The related walking processes can be optically monitored in real time. Also, we have demonstrated that the walker number and the optical response of the system can be correlated. Our studies may outline a very important step toward the development of advanced artificial nanomachinery[43,44] with multiple walkers working in concert. From the viewpoint of plasmonics, our studies also offer a solution to deterministic placement and dynamic manipulation of individual nanoparticles with nanoscale precision. This may enable active plasmonic devices with prescribed levels of performance and functionalities, through controlled motion and interaction among multiple plasmonic nanoparticles.

FIGURES

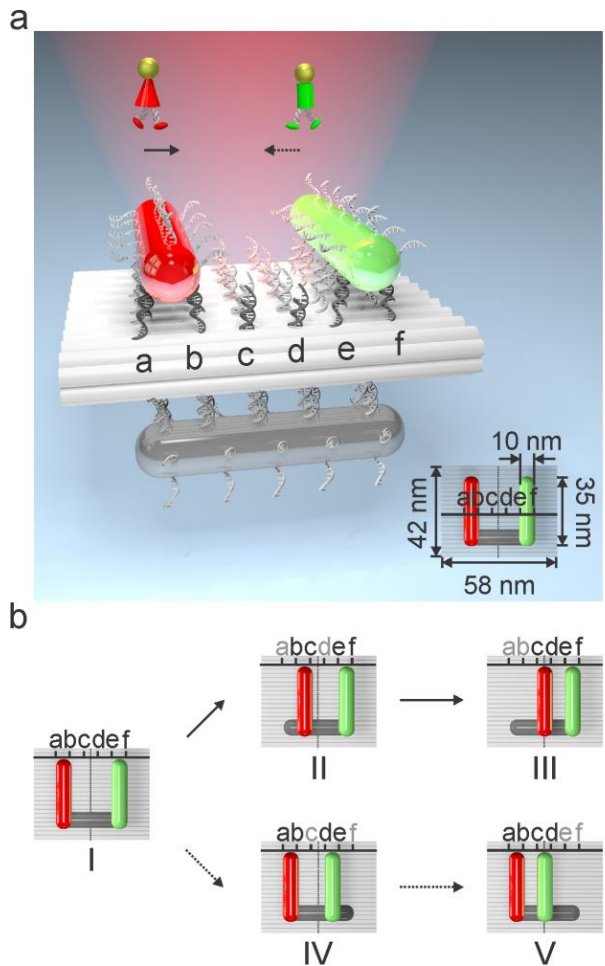

**Figure 1.** (a) Schematic of the plasmonic walker couple. The red beam indicates the incident circularly polarized light. (b) Stepwise walking of the individual walkers. At state I, the walkers are symmetrically positioned at the two ends of the stator. Along the upper route, the green walker halts and the red walker takes two successive steps, reaching states II and III, respectively. Along the lower route, the red walker halts and the green walker carry out two successive steps, reaching states IV and V, respectively.

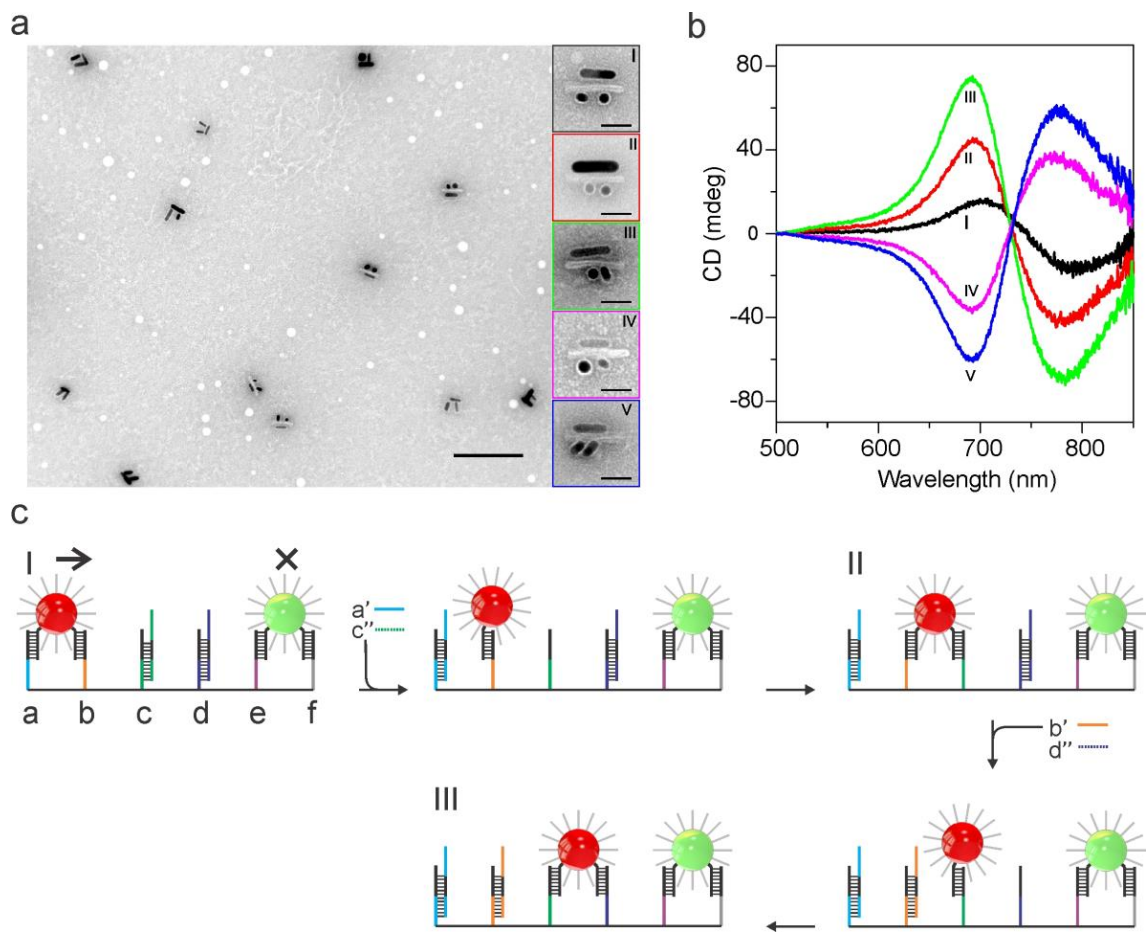

**Figure 2.** (a) Overview TEM image of the plasmonic walker couple structures at state I. Scale bar: 200 nm. Magnified views: structures at states I-V. Scale bar: 30 nm. The plasmonic structures display certain deformation on the TEM grid. (b) Measured CD spectra at different states. (c) Schematic of the walking process from state I to state III. By adding blocking and removal strands, the red walker programmably attaches (detaches) its feet on (from) the track through hybridization (de-hybridization) with the corresponding foothold rows.

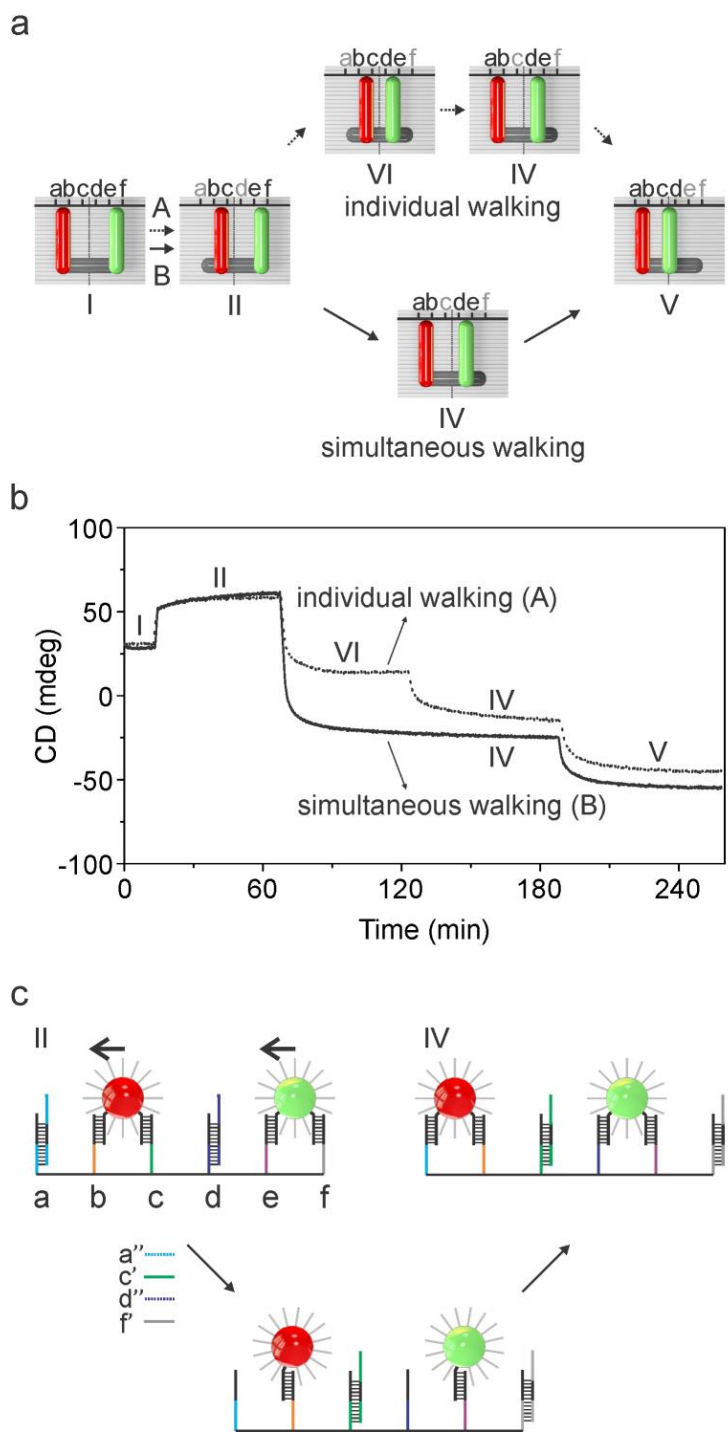

**Figure 3**. (a) Routes for individual (I-II-VI-IV-V, route A) and simultaneous (I-II-IV-V, route B) walking. (b) Dynamic processes of individual and simultaneous walking detected by in-situ CD spectroscopy at a fixed wavelength of 695 nm. (c) Schematic of the simultaneous walking process.

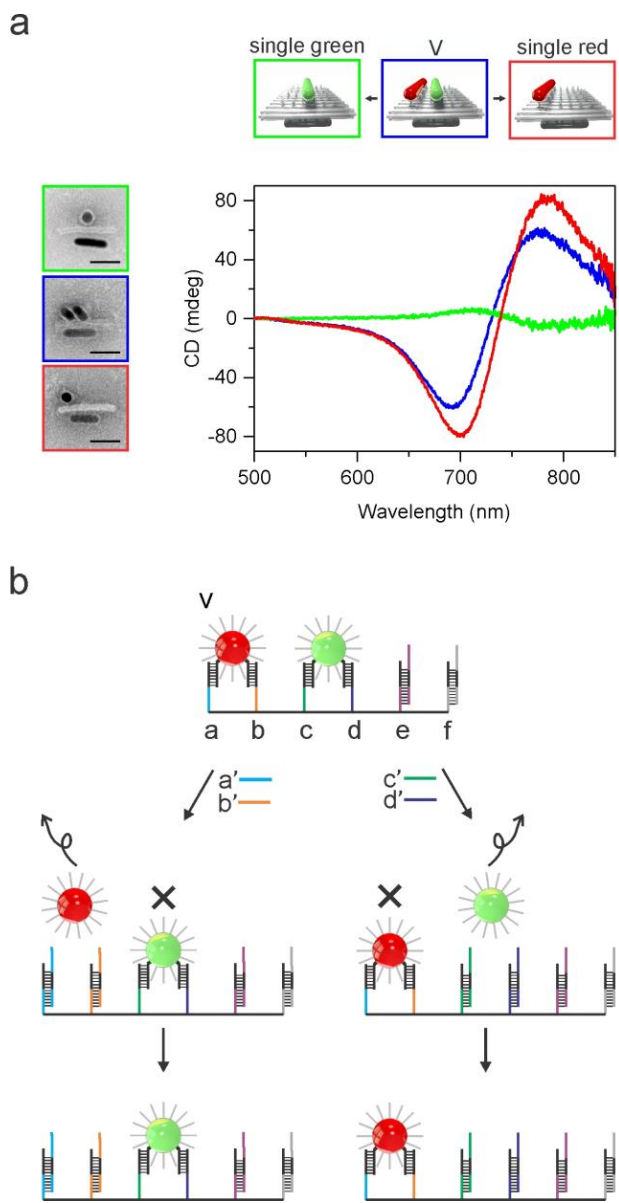

**Figure 4.** (a) TEM images and measured CD spectra of the plasmonic walker couple structures at state V and the structures after selective dissociation of one walker off the track. Scale bar: 30 nm. (b) Schematic of the programmable dissociation process.

**Supporting Information**. Materials and methods, theoretical calculations, additional data and detailed walking procedures. This material is available free of charge via the Internet at http://pubs.acs.org.

AUTHOR INFORMATION

**Corresponding Author**
*E-mail: laura.liu@is.mpg.de

**Author Contributions**

M.U., C.Z., and N.L. designed the project, analyzed the data, and wrote the manuscript. M.U. and C.Z. performed the experiments. X.D. did the theoretical calculations.

**Notes**

The authors declare no competing financial interest.

**Acknowledgements**

We thank A. Jeltsch and R. Jurkowska for assistance with CD spectroscopy. We thank M. Kelsch and K. Hahn for assistance with TEM microscopy. We thank M. Hentschel and A. Kuzyk for comments on the manuscript. TEM data was collected at the Stuttgart Center for Electron Microscopy (StEM). This project was supported by the Sofja Kovalevskaja grant from the Alexander von Humboldt-Foundation, the Marie Curie CIG grant, and the European Research Council (ERC *Dynamic Nano*) grant.

**Table of Contents**

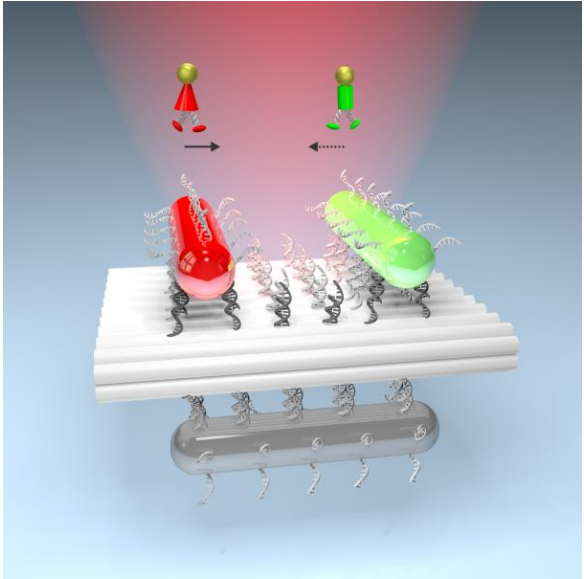